\documentclass[pra,preprint,tightenlines,showpacs,twocolumn,10pt]{revtex4}\usepackage{epsfig} 
     
\begin{document} 

\title{ A new mechanism for electron transfer in fast ion-atomic collisions } 

\author{ A.B.Voitkiv, B.Najjari and J.Ullrich } 
\affiliation{ Max-Planck-Institut f\"ur Kernphysik, 
Saupfercheckweg 1, D-69117 Heidelberg, Germany} 

\begin{abstract} 
We discuss a new mechanism for the electron capture 
in fast ion-atom collisions. Similarly like in the radiative capture,  
where the electron transfer occurs due to photon emission, 
within the mechanism under consideration the electron capture 
takes place due to the emission of an additional electron. 
This first order capture process  
leads to the so called transfer-ionization 
and has a number of interesting features, 
in particular, in the target frame 
it results in the electron emission 
mainly into the backward semi-sphere. 

\end{abstract} 

\pacs{34.10.+x, 34.50.Fa} 

\maketitle 


When a projectile-ion collides with an atom 
an electron can be transferred from 
the atom to the ion. Such a transfer process, 
which is also termed electron capture, 
represents one of the basic processes considered 
in atomic physics. During the last three decades 
the electron capture in fast ion-atom collisions 
has been the subject  of very intense research  
(see e.g. \cite{review} 
where further references can be found).
 
There are several basic mechanisms 
for the electron capture which 
have been considered so far. 
They can be divided 
into the radiative and nonradiative ones. 

Amongst the nonradiative capture channels 
the largest cross sections are, as a rule,  
yielded by the so called kinematic capture. 
This capture mechanism, first considered 
by Oppenheimer and Brinkman and Kramers \cite{OBK}, 
appears already in first order perturbation theory.  
  
An interesting peculiarity of the nonradiative 
capture is that at asymptotically high values of the impact 
velocity $v$ it would be dominated 
by a second order - the so called Thomas  - transfer mechanism. 
Within the latter, initially proposed in \cite{thomas}  
within the framework of classical mechanics,  
the capture proceeds in two steps. 
First, as a result of a binary collision with 
the projectile the electron acquires a velocity whose 
absolute value is equal to that of the incident projectile. 
Second, after a subsequent collision with 
the target nucleus the electron emerges 
with a velocity vector equal to that of 
the projectile that makes the capture very probable.  
If the target has more than one electron then, 
after a hard collision with the projectile, 
the electron could scatter on another target electron 
acquiring the velocity of the projectile. These two 
second order mechanisms are often referred 
to as the (target) nucleus-electron Thomas (NET) 
and the electron-electron Thomas (EET) mechanisms, 
respectively. Using quantum mechanics the NET and EET 
were considered, for the first time, in \cite{drisko} and 
\cite{briggs}, respectively. 

In the radiative capture mechanism the electron transfer 
occurs due to the emission of a photon with a frequency 
$\sim v^2/2$ \cite{b-d}. 
Like the kinematic capture, the radiative capture  
channel appears already in first order perturbation 
theory also representing a first order transfer 
mechanism. 

In the present communication we shall consider 
yet another first order capture mechanism. It  
becomes possible when the target has initially 
more than one electron, bears important 
similarities to the kinematic and radiative captures 
but also has a number of interesting features.  

At nonrelativistic impact energies the total 
cross sections, both for the nonradiative 
and radiative capture, are 
Galilean invariant \cite{voit-galilei-gauge}.  
Nevertheless, a proper choice 
of the reference frame is often of crucial importance 
leading to drastic simplifications and enabling 
one to understand much easier the basic physics 
of the process. 

For example, if we consider the radiative capture 
using the rest frame of the target atom, 
it is not that easy to understand 
why the emission of a photon with an energy $\omega \sim v^2/2$ 
by the electron (which is going to 'jump' into a bound state 
of a fast moving ion and, thus, must increase its own energy 
by about the same amount $ \sim v^2/2$) 
helps the electron transfer to occur. 
However, the same transfer process becomes 
very transparent when analyzed in the projectile frame.  
Indeed, in this frame the electron has initially 
a kinetic energy $v^2/2$ (which is the initial 
internal kinetic energy of the subsystem 
'projectile+electron') and, in order   
to undergo the transfer, must rid of 
the energy excess $\sim v^2/2$ which 
occurs via the photon emission. 

\begin{figure}[t] 
\vspace{-0.45cm}
\begin{center}
\includegraphics[width=0.37\textwidth]{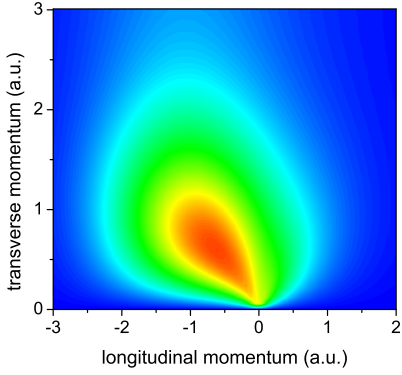}
\end{center}
\vspace{-0.9cm} 
\caption{ \footnotesize{ Momentum spectrum of 
the recoil electron emitted via the E-E capture   
in $1$ MeV p$^+$ + He(1s$^2$) $\to$ H(1s) + He$^2$ + e$^-$ 
collisions. The spectrum is given in the 
rest frame of the target atom.}}  
\label{figure1} 
\end{figure}

Let us now assume that the atom has two electrons. 
Considering the possibilities for the electron capture and 
working again in the rest frame of the projectile 
we see that now the electron to be transferred 
can try to rid of the energy excess not only 
via the coupling to the radiation field 
but also by the interaction with the other electron. 
Since the energy excess is large (on the atomic scale)  
one can expect that the outcome of such an energy 
transfer will be emission of the 'recoil' electron. 
In the projectile frame the recoil electron 
will be emitted with a velocity larger than 
the atomic velocity $v$ and intuitively 
it is also rather clear that this emission 
will proceed mainly in the direction 
of the atomic motion in this frame. 
The latter implies that in the rest frame of 
the atom the recoil electron  
will mainly be emitted in the direction 
opposite to the projectile motion,  
i.e. into the backward semi-sphere (see for illustration 
figure \ref{figure1}). 

Similarly to the case of the radiative capture, 
the simplest treatment for this electron-electron 
(E-E) capture mechanism is given if we consider 
the process in the rest frame of the projectile 
and start with the approximate first order 
transition amplitude, which  reads 
\begin{eqnarray}
a_{fi} = -i \int_{-\infty}^{+\infty} dt %
\langle \Psi_f(t) |\hat{W} |\Psi_i(t)\rangle.       %
\label{e1} 
\end{eqnarray}  
Here $\hat{W}=1/\mid {\bf r}_1 - {\bf r}_2 \mid  $
is the interaction between the electrons 
whose coordinates in the projectile frame are 
${\bf r}_1$ and ${\bf r}_2$ and   
the initial and final two-electron states 
are approximated by  
\begin{eqnarray}
\Psi_i(t) &=& \varphi_i({\bf s}_1,{\bf s}_2) %
e^{i {\bf v} \cdot ({\bf r}_1+{\bf r}_2)} %
e^{-i (\epsilon_i + v^2 ) t} %
\nonumber \\  
\Psi_f(t) &=&  \hat{A} %
\chi_f({\bf r}_1) \phi_f({\bf s}_2) %
e^{-i (\varepsilon_f + \epsilon_f) t} %
e^{i {\bf v} \cdot {\bf r}_2 }  
e^{-i v^2/2 t }.     
\label{e2} 
\end{eqnarray}  
In Eqs.(\ref{e2}) 
${\bf s}_j={\bf r}_j-{\bf R}(t)$ ($j=1,2$), 
${\bf R}(t)= {\bf b} + {\bf v} t$ is 
the classical straight-line trajectory of the atomic nucleus,  
$\varphi_i$ is the initial (unperturbed) two-electron 
atomic state with an energy $\epsilon_i$, 
$\chi_f$ is the (unperturbed) bound state 
of the captured electron, $\phi_f$ is the  
state of the recoil electron, 
$ e^{i {\bf v} \cdot {\bf r}_j -i v^2/2 t}$ is 
the so called translational factor, 
$\hat{A}$ is the (anti)symmetrization 
electron operator and it is assumed 
that the recoil electron moves 
in the field of the residual atomic 
ion and $\epsilon_f$ is its 
final energy (in the atomic frame). 

For the helium target the initial state 
was taken as 
\begin{eqnarray}
\varphi_i({\bf s}_1,{\bf s}_2) = %
A_i \left( e^{-\alpha s_1 - \beta s_2} %
+ e^{-\alpha s_2 - \beta s_1} \right) %
e^{\gamma s_{12}},     
\label{in-state}  
\end{eqnarray}  
where $A_i$ is the normalization 
and the following sets of the parameters 
$\alpha$, $\beta$ and $\gamma$ were considered: 
(i) $\alpha=\beta=1.69$, 
$\gamma=0$; (ii) $\alpha=\beta=1.86$,  
$\gamma=0.254$; (iii) $\alpha=2.18$, $\beta=1.18$,  
$\gamma=0$; and (iv) $\alpha=2.21$, $\beta=1.44$,  
$\gamma=0.207$.  
 
Note that from the perspective, chosen here 
to consider the electron capture, there is not only 
the apparent similarity between the E-E 
and radiative captures but also the kinematic capture 
bears an important resemblance to the former two. 
Indeed, viewing the latter process in the rest frame 
of the projectile the kinematic capture can be interpreted 
as occurring due to the interaction between the electron 
and the atomic nucleus during which the energy 
excess $\sim v^2/2$ is transferred to this nucleus. 
In a more formal consideration this simple picture 
follows especially naturally if the kinematic capture 
is considered using the post form of 
the Oppenheimer-Brinkman-Kramers (OBK) approximation. 

Thus, the kinematic, E-E and radiative capture 
channels have important common points. 
In particular, all of them are 
first order mechanisms and the role of the projectile 
in these capture processes is not to 'hijack'  
the electron via a violent interaction but rather 
to merely offer additional 'seats' for it. 
The latter makes the atomic configuration unstable 
and allows the interaction between the particles 
of the atom (and/or with the radiation field) 
to break the atom which, in the absence of these 'seats',  
would be perfectly stable.      

The theoretical consideration of 
the E-E capture, based on the approximate 
states (\ref{e2}), corresponds to 
the description of the radiative capture 
within the so called first-Born (1B) 
approximation (see e.g. \cite{croth})    
and to the treatment of the kinematic capture 
within the OBK approximation. Therefore, we first 
compare the E-E, kinematic and radiative 
capture cross sections obtained within these 
simplest approaches. 

Analysis shows that the dependences 
of E-E capture cross section 
on the projectile charge $Z_p$ 
and collision velocity $v$ are similar 
to those of the kinematic capture. 
In particular, in the asymptotic limit 
$v \gg Z_p, Z_t$ ($v \to \infty$), 
where $Z_t$ is the charge of the target nucleus, 
the total E-E capture cross section is roughly 
proportional to $Z_p^5/v^{12} $. 
However, the dependence on the charge $Z_t$ 
(in the case of a two-electron target) 
for the E-E capture is roughly $\sim Z_t^3$ 
which is by about a factor of $Z_t^2$ weaker 
compared to the dependence $\sim Z_t^5$ 
for the kinematic capture. 

This difference can be understood 
by noting that the interaction between the target 
electron and the target nucleus in 
the OBK amplitude, given by $- Z_t/r$, 
and the interaction $1/r_{12}$ 
between the target electrons 
in the amplitude (\ref{e1}) 
scale differently with $Z_t$. 
While the electron-electron interaction 
increases with $Z_t$ only because the size 
of the atomic system (and, thus, $r_{12}$) 
decreases, the electron-nucleus interaction 
$-Z_t/r$ increases also because 
of the factor $Z_t$ which eventually results 
in an 'extra' factor $\sim Z_t^2$ 
in the OBK capture cross section.    

Note also that in case of multi-electron 
targets the electron to be transferred 
has many 'partner'-electrons  
to interact with. Because of the cumulative 
effect of such interactions on 
the E-E capture channel, the dependence 
of the E-E capture cross section 
on $Z_t$ may become somewhat 
stronger compared 
to the case of two-electron systems.   
 
Further, comparing the E-E capture 
with the radiative one 
we see that the dependence on the projectile 
charge $Z_p$ for these two mechanisms 
is quite similar. However, 
the E-E capture cross section 
decreases much faster 
with the collision velocity and, 
in contrast to the radiative  
capture, depends (per target electron) 
on $Z_t$. The latter difference   
arises because the electron-electron interaction 
depends on $Z_t$ while the interaction between 
the electron and the radiation field at 
$v \gg Z_t$ is practically $Z_t$-independent. 
Concerning the different dependences 
on the collision velocity, it can be  
traced back to originate from the different 
dispersion relations between the energy and momentum 
inherent to a nonrelativistic electron 
and a photon \cite{f2}.   
  
Equal masses of the captured and recoiling electrons  
also lead to the fact that, in contrast to the 
radiative capture, within the E-E capture mechanism 
not the full amount $\sim v^2/2$ of the energy excess 
is taken by the recoil electron: in the projectile frame 
the incident atomic nucleus also gets a part of 
this energy and this part, for a given value of $v$, 
increases with $Z_t$. 
It, however, is noticeably smaller than 
the part of the energy excess 'absorbed' 
by the recoil electron and it tends 
to decrease (in relative terms) 
when the collision velocity $v$ increases.     

In our calculations of the E-E capture 
from helium we used all four options (i)-(iv) to approximate 
the initial state (\ref{in-state}). 
While we observed a certain sensitivity 
of the total E-E capture cross section  
to the choice of $\alpha$, $\beta$ and $\gamma$ 
(variations of up to $40 \%$), 
all these options yielded a very similar emission 
pattern for the recoil electron. 

\vspace{0.15cm} 
   
Up to now we were discussing the capture processes using 
the simplest theoretical treatments. It is well known, 
however, that the values of the kinematic capture 
cross section are poorly described by the OBK approximation, 
even at very high impact energies \cite{review}.   
The 1B model for the radiative capture 
also does not represent the best description 
of this process. Therefore, one can expect 
that the simple treatment of the E-E capture, 
discussed above, also needs to be substantially refined. 

The continuum-distorted-wave model (CDW) 
is known to represent a great improvement  
of the OBK model and has been successfully 
applied to compute the total cross sections 
for the nonradiative capture \cite{review}. 
Within this model the initial and final 
electron wave functions are approximated  
by two-center states.  
In the case of electron capture occurring 
in a three-body coulomb system 
of colliding bare projectile-nucleus 
and a hydrogen-like atom/ion 
the initial and final states of 
the transferred electron, treated  
in the rest frame of the projectile,  
are given by   
\begin{eqnarray}
\psi_i^{(+)}(t) &=& L_i^{(+)}({\bf r}) \, %
\varphi_i({\bf r}-{\bf R}) %
e^{i {\bf v} \cdot {\bf r} } %
e^{-i (\epsilon_i + v^2/2 ) t }, %
\nonumber \\ 
\psi_f^{(-)}(t) &=& L_f^{(-)}({\bf r}-{\bf R}) \, \chi_f({\bf r}) %
e^{-i \varepsilon_f t}.  %
\label{e4} 
\end{eqnarray} 
In Eq.(\ref{e4}) $\varphi_i$ and 
$\chi_f$ are, respectively, 
the initial and final undistorted 
bound states of the electron. 
The factors $L_i^{(+)}$ and 
$L_f^{(-)}$  
account for the distortions of these states 
caused by the field of the projectile 
(in the initial channel) and the field 
of the residual atomic ion 
(in the final channel). 
These factors read   
\begin{eqnarray}
 L_i^{(+)}({\bf r}) &=& 
 N(\nu_p) \, \left._1F_1 \right. %
\left( i\nu_p,1,i v r- i {\bf v} \cdot {\bf r} \right) %
\nonumber \\ 
L_f^{(-)}({\bf s}) &=& N^*(\nu_t) \, %
\left._1F_1 \right. %
\left( -i\nu_t,1, - i v s - i {\bf v} \cdot {\bf s} \right),  
\label{e5} 
\end{eqnarray} 
where $ N(\nu)= e^{ \pi \nu /2 } \Gamma(1-i\nu)$, 
$\nu_p=Z_p/v$, $\nu_t=Z_t/v$, 
and $\Gamma$ and $ \left._1F_1 \right.$  
are the gamma and confluent 
hypergeometric functions, 
respectively.  
The CDW model for the nonrelativistic  
radiative capture \cite{croth} uses the same 
distortion factors and it has demonstrated 
a great advantage \cite{voit-galilei-gauge} 
over the 1B model.  

Therefore, in the spirit of the CDW approach  
we have attempted to improve the consideration 
of the E-E capture channel by incorporating 
the distortion factors (\ref{e5}) into the states 
(\ref{e2}). 
Note, however, that such a change 
in the description, while leading 
to a substantial reduction  
of the E-E capture cross section values, 
turned out to have a very little influence   
on the shape of the emission pattern 
of the recoil electron and 
the energy/momentum sharing 
between the recoil electron 
and the atomic nucleus. 
Besides, concerning 
the dependences on the charges 
$Z_p$ and $Z_t$ as well as on 
the collision velocity $v$,  
the correspondence between the E-E   
and the kinematic capture channels, 
both treated using the distortion factors,    
remain practically the same as 
in the simple models discussed above:  
at asymptotically high $v$  
these two channels show similar dependences on 
$Z_p$ and $v$ and again the E-E capture channel 
demonstrates the less steeper increase with $Z_t$.   

Our calculations show that the E-E capture channel, 
while easily 'beating' the radiative capture  
at not too large collision 
velocities, yields cross section values which 
always represent just a fraction (a few per cent) 
of those of the kinematic capture. However, 
despite this relative weakness, 
the E-E channel remains of importance 
since its leads to the transfer-ionization and thus 
to a different charge state of the residual target ion. 
Therefore, it is more appropriate to compare the 
E-E channel with other reaction channels leading to 
the transfer-ionization. 

The direct counterpart of the E-E capture 
is the EET mechanisms. At sufficiently high impact 
velocities the latter would dominate the E-E mechanism 
since the EET, as a second order capture process, 
weakens more slowly with increasing $v$. 
However, according to the results of \cite{briggs}, 
at collision velocities of $\sim 10$-$20$ a.u. 
the ratio of the EET to the kinematic capture  
is just about $0.1$-$0.2 \%$. Therefore, 
at these impact velocities the contribution 
of the EET to the transfer-ionization 
is at least by about an order of magnitude 
smaller than that of the E-E mechanism.    

Other 'rivals' of the E-E capture channel  
are the kinematic capture accompanied by 
shake-off of the second electron (KEC-SO) 
and the independent transfer-ionization (ITI). 
In the former, due to the fast removal 
of the captured electron, the other 
atomic electron 'sees' a sudden change 
in the atomic potential and 
has a nonzero probability 
to become unbound \cite{mcg}.  
In the ITI the electrons make the corresponding 
transitions independently of each other.    
This process appears in second order perturbation 
theory and for its realization does not need  
any electron-electron interaction.  
One of signatures of these two processes is that 
the momentum distributions of the recoil ions 
are centered at the positions characteristic for 
the (single) kinematic capture.  

\begin{figure}[t] 
\vspace{-0.45cm}
\begin{center}
\includegraphics[width=0.49\textwidth]{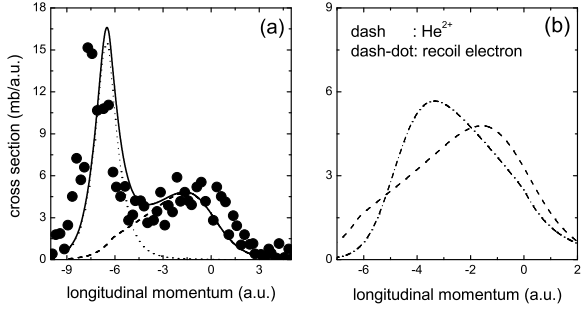}
\end{center}
\vspace{-0.9cm} 
\caption{ \footnotesize{ Longitudinal momentum 
spectrum of the electrons and He$^{2+}$ 
in the transfer-ionization of He(1s$^2$) by 
$4$ MeV p$^+$. (a) Solid curve: 
the calculated spectrum of He$^{2+}$. 
Dash and dot curves: the contributions 
of the E-E and the two-step and shake-off channels, respectively. 
Symbols: experimental data from \cite{schmidt} 
(normalized to the total cross section 
from table I of \cite{schmidt}).  
(b) The recoil electron and ion spectra  
produced via the E-E mechanism. }}  
\label{figure2} 
\end{figure}

The transfer-ionization in proton-helium collisions 
has attracted recently much attention \cite{mergel}-\cite{schmidt}  
triggered by the experimental observation \cite{mergel}  
that a very substantial part of the emission is formed 
by the electrons ejected in the backward direction. 
This feature prompted the authors of \cite{mergel} 
to speculate that it might be an indication 
of certain features of the electron correlation   
in the initial ground state of the helium target. 
This hypothesis was theoretically studied in \cite{frank}. 

In this respect it is worth noting that 
the backward emission is a natural signature of 
the E-E capture mechanism (see figure \ref{figure1}). 
Concerning the electron-electron correlation/interaction 
in the ground state, one can note that 
it increases the distance between the electrons and  
effectively weakens the E-E capture channel: for example, 
if we fully neglect the influence 
of the electron-electron interaction 
on the ground state by setting 
$\alpha=\beta=2$ and $\gamma=0$ the calculated 
E-E capture cross section becomes much larger 
(but the emission pattern remains basically the same). 
   
In figure \ref{figure2} we show  
the longitudinal spectra of 
the emitted electrons and target recoil 
ions produced in $4$ MeV p$^+$ + He(1s$^2$) 
$\to$ H(1s) + He$^{2+}$ + e$^-$ 
collisions ($v=12.7$ a.u.). 
The ion spectrum consists of two peaks. 
The right broad peak appears due to the E-E mechanism.  
The left one has its origin mainly in the 
KEC-SO process; besides, the ITI process gives 
a noticeable contribution to this peak.   
The E-E contribution was calculated using 
the wave function (\ref{in-state})-iv   
and including the distortion factors. 
Details of our calculations shown in figure  
\ref{figure2} will be given in a forthcoming 
paper. 

The authors of \cite{schmidt},   
based on the approximate 
scaling $\sim v^{-11}$ for their measured 
transfer-ionization cross sections,   
interpreted the right peak as caused by the EET mechanism. 
However, at the impact velocities 
$v \simeq 8$-$15$ a.u. considered in \cite{schmidt}, 
the asymptotic regime is not yet reached:     
the cross sections, both for the kinematic and E-E captures, 
also behave approximately as $ v^{-11}$ and 
the E-E mechanism still strongly dominates over the ETT. 

Compared to the case with protons  
the capture cross sections very strongly 
increase in collisions with multiply charged ions. 
Although the transfer-ionization process 
in such collisions will be dominated 
by the ITI process, 
the contribution of the E-E channel 
may still be separated because 
of its characteristic 
backward emission pattern.  

In summary, we have considered 
a capture mechanism in fast ion-atom collisions 
which was not discussed so far. Within this mechanism 
an atomic electron is captured via the interaction with another 
atomic electron(s) which plays a role similar to that of 
the interaction with the radiation field 
in the radiative capture. 
Alike the radiative and kinematic capture channels 
the E-E capture represents a first order transfer mechanism. 
However, in contrast to the former two, the latter 
leads mainly to the transfer-ionization. 
A prominent feature of the E-E capture  
is the backward emission of the recoil electrons  
which makes this process visible even 
if the total cross section is strongly 
dominated by other capture channels. 
The E-E mechanism could 
also be of importance for charge states 
of fast ions under channeling in crystals where 
the kinematic capture is strongly suppressed.


\begin{thebibliography}{99}
\bibitem{f2} 
In a system containing heavy and 
light 'electrons' the capture 
of a 'heavy electron' accompanied 
by the emission of a 'light electron' 
would be quite similar to the radiative capture.   
Such a situation might arise in 
nuclear collisions when one of 
the collision partners is 
a neutral atom and the transfer 
of a proton (a 'heavy electron') 
between the nuclei could proceed 
with emission of an atomic electron 
(a 'light electron').
\end{thebibliography}
\end{document}